\documentclass[prb,twocolumn,amsmath,showpacs]{revtex4}

\usepackage{epsfig}
\usepackage{graphicx}

\begin{document}
\input epsf

\title { Comparison of the scaling analysis of the mixed-state magnetization data with direct measurements of the upper critical field in YBa$_2$Cu$_3$O$_{7-x}$}

\author {I. L. Landau}
\affiliation{University of Berne, Department of Chemistry and Biochemistry, Freiestrasse 3, CH-3012 Berne, Switzerland}
\affiliation{Kapitza Institute for Physical Problems, 117334 Moscow, Russia}

\date{\today}

\begin{abstract}

By comparison of recent direct measurements of the temperature dependence of the upper critical field $H_{c2}$ in an YBa$_2$Cu$_3$O$_{7-x}$ high-$T_c$ superconductor with the scaling analysis of magnetization data, collected in fields $H \ll H_{c2}$, we demonstrate that that the temperature dependence of the Ginzburg-Landau parameter $\kappa$ is negligible. Another conclusion is that the normalized temperature dependence of $H_{c2}$ is independent of the orientation of the magnetic field in respect to crystallographic axes of the sample. We also discuss that isotropy of the temperature dependence of $H_{c2}$ straightforwardly follows from the Ginzburg-Landau theory if $\kappa$ does not depend on temperature.

\end{abstract}

\maketitle

Evaluation of the upper critical field $H_{c2}$ and its temperature dependence represents a difficult task if high-$T_c$ superconductors (HTSC) are concerned. The problem is that $H_{c2}$ is very high and can be measured directly only in pulsed magnetic fields of megagauss amplitudes. This is an obvious reason that only several such studies were published so far and not all of them may be considered as reliable measurements. We could find only a very few works, in which measurements were extended to a considerable range of $T/T_c$ and all of them were made on YBa$_2$Cu$_3$O$_{7-x}$ samples.\cite{nakao-jp,nakao-pb,miura,sekitani-pb,sekitani}

At the same, $H_{c2}$ represents one of the main parameters of a superconductor and its knowledge is of primary importance. This is why several indirect approaches have been proposed and used in order to evaluate $H_{c2}(T)$ from equilibrium magnetization data collected in fields $H \ll H_{c2}$.\cite{h-clem1,h-clem2,blk0,blk1,blk2,blk3,land1} However, all these approaches are based on certain assumptions, which are not necessarily satisfied in experiments. This makes existing $H_{c2}(T)$ results questionable. 

We shall not consider all theoretical methods for the analysis of magnetization data. Our goal is to discuss a scaling procedure, proposed in Ref. \onlinecite{land1}, in order to compare the normalized temperature dependencies of $H_{c2}$, obtained by employing this procedure, with direct measurements of the upper critical field. As we demonstrate below, good agreement between the results provides convincing evidence of the validity of this scaling analysis and allows to make some conclusions about the temperature dependence of the Ginzburg-Landau parameter $\kappa$.. 

The scaling procedure is based on a single assumption that $\kappa$ is temperature independent. In this case, equilibrium mixed state magnetizations measured at different temperatures but in the same normalized fields $H/H_{c2}(T)$ are proportional to $H_{c2}(T)$. This is true in fields $H \gg H_{c1}$, i.e., this situation can only be achieved in high $\kappa$ superconductors, which is the case for HTSC's as well as for many other novel superconducting materials.

According to Ref. \onlinecite{land1}, the magnetizations of a sample at two different temperatures $T$ and $T_0$ are related by
\begin{equation}
M(H,T_0)=M(h_{c2}H,T)/h_{c2}+c_0(T)H,
\end{equation}
where $h_{c2} = H_{c2}(T)/H_{c2}(T_{0})$ is the normalized uper critical field and $c_{0}(T)= \chi_{n}(T_{0}) - \chi_{n}(T)$ ($\chi_{n}$ is the normal-state magnetic susceptibility of a sample). The first term on the right side of Eq. (1) describes the properties of the mixed state of ideal type-II superconductors, while the second one is introduced in order to account for all other temperature dependent contributions to magnetization, which are unavoidable in HTSC's. 

By a suitable choice of $h_{c2}$ and $c_{0}(T)$ individual $M(H)$ curves measured at different temperatures may be merged into a single master curve $M_{eff}(H,T_0)$. In this way the temperature dependence of the normalized upper critical field is obtained.\cite{land1} 

This approach turned out to be quite effective for the analysis of reversible magnetization data.\cite{land1,land2,land3,land4,land5, thomp,doria,NbSe,LiPdB} The main and unexpected result of these analyses is that all numerous HTSC's may be divided into two groups. The dependencies of the normalized upper critical field  $H_{c2}$ on $T/T_c$ for HTSCÕs belonging to the same group are practically identical, while they are distinctly different between the groups. The larger group includes a huge variety of various HTSC compounds, while the second one is rather small and consists of just several cuprates.\cite{land1,land2,land3,land4}, Apparently, a level of doping plays an important role in this matter.\cite{land1,land2} In the following, we shall discuss only the larger group of HTSC's because the corresponding $H_{c2}(T)$ curve is indeed close to results of Refs. \onlinecite{nakao-jp,nakao-pb,miura,sekitani-pb,sekitani}.

Another important result of the scaling analysis is that practically the same curves were obtained for single crystals, grain-aligned samples, and ceramics.\cite{land2,bern} This indicates that, in spite of strong anisotropy of absolute values of $H_{c2}$, its normalized temperature variation depends very little on the orientation of an applied magnetic field. In other words, if the temperature dependence of the upper critical field is written as
\begin{equation}
H_{c2}(T) = H_{c2}(0)F(1 - T/T_c),
\end{equation}
$H_{c2}(0)$ depends on the orientation of the magnetic field, while the function $F$ is isotropic.

As was argued in Ref. \onlinecite{land1}, the fact that the analyses of magnetization data for many different HTSC's result in practically identical normalized $H_{c2}(T/T_c)$ cannot be a coincidence but rather represents strong evidence that this approach is generally correct. At the same time, it does not necessary mean that $\kappa$ is temperature independent. Indeed, the universality of $H_{c2}(T/T_c)$ will not be altered if $\kappa$ is temperature dependent, but this dependence is the same for different HTSC's. 

If paramagnetic contribution to the sample magnetization is negligible or it can be evaluated with sufficient accuracy, there will remain only one adjustable parameter in Eq. (1) and, in this case, the temperature dependence of $\kappa$ can also be evaluated from the scaling analysis of magnetization data, as it was demonstrated in experiments with low-$T_c$ superconductors.\cite{NbSe,LiPdB} However, this is not the case for HTSC's, in which the normal-state paramagnetic contribution is always substantial and can hardly be evaluated independently. In other words, the main assumption about temperature independence of $\kappa$ in HTSC's  has never been tested. 

This is why, we consider recent direct measurement of $H_{c2}(T/T_c)$ in pulsed magnetic fields\cite{sekitani-pb,sekitani} as a unique opportunity for such a test. An important advantage of these works is that a  new method of a radio frequency transmission was developed. This technique allows for evaluation of $H_{c2}$ with substantially better accuracy than previously used magnetoresistance measurements. 

It seems to be commonly accepted that both $H_{c2}(0)$ and $F$ in Eq. (2) depend on the orientation of the magnetic field.\cite{sekitani-pb,sekitani} We could not, however, find any experimental confirmations of this in the literature. Furthermore, it seems that $H^{(c)}_{c2}(T)/H^{(c)}_{c2}(0)$ and $H^{(ab)}_{c2}(T)/H^{(ab)}_{c2}(0)$ are practically identical, i.e., $F$-function in Eq. (2) is indeed isotropic as it was argued on the basis of the scaling analysis. \cite{land2}

\begin{figure}[h]
 \begin{center}
  \epsfxsize=1\columnwidth \epsfbox {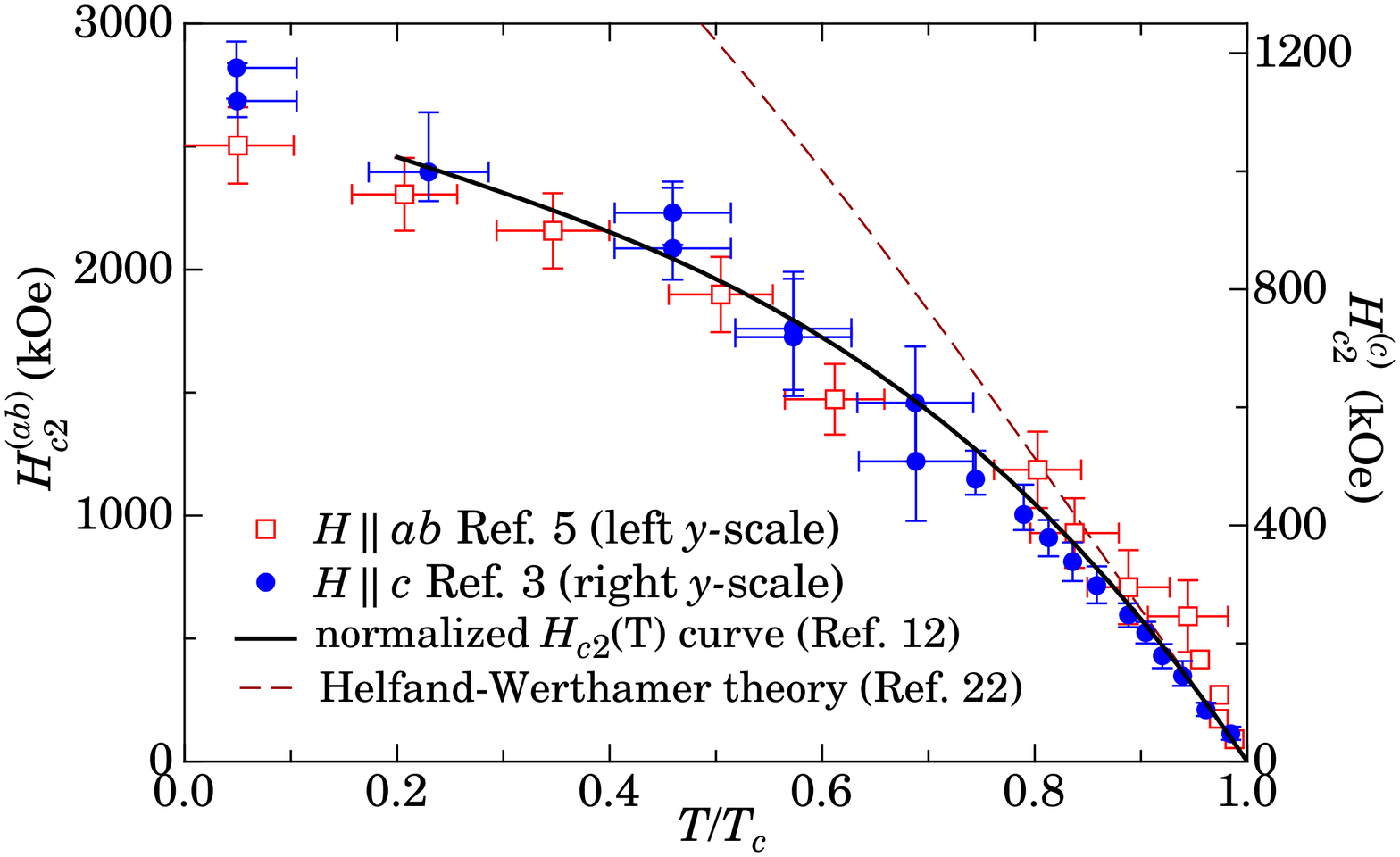}
  \caption{(Color online) Experimental $H^{(ab)}_{c2}(T/T_c)$ and $H^{(c)}_{c2}(T/T_c)$ data from Refs. \onlinecite{miura} and  \onlinecite{sekitani}, respectively. The solid line is the normalized $H_{c2}(T/T_c)$ curve obtained by scaling of equilibrium magnetization data and  fitted to data points. }
 \end{center}
\end{figure}
In Fig. 1 we plot $H^{(ab)}_{c2}(T/T_c)$  (left $y$-scale) and $H^{(c)}_{c2}(T/T_c)$ (right $y$-scale) from Refs. \onlinecite{sekitani} and  \onlinecite{miura}, respectively. The latter results were obtained by magnetoresistance measurements carried out on epitaxial YBa$_2$Cu$_3$O$_{7-x}$ film. The zero-field resistive transition was about 4 K wide with the zero resistance state below 83.5 K. This provides considerable uncertainty in $T_c$. For this plot $T_c$ was chosen by  extrapolation of the $H_{c2}(T)$ curve, presented in Fig. 4 of Ref. \onlinecite{miura}, to $H_{c2}(T) = 0$. This gives $T_c = 87.5$ K, which is the very upper end of the resistive transition. As may be seen in Fig. 1, both data sets match each other quite well and the difference between them does not exceed uncertainty of the results. 

The normalized temperature dependence of $H_{c2}$, obtained by scaling of magnetization data in fields $H \ll H_{c2}$,\cite{land1} which is also shown in Fig. 1, perfectly fits experimental data. This agreement with direct experimental results strongly supports the main assumption about temperature independence of $\kappa$. Indeed, even a rather weak temperature dependence of $\kappa$, predicted by the conventional BCS theory,\cite{wert} changes the resulting $h_{c2}(T/T_c)$ curve in a way that it cannot satisfactory describe experimental data (see Ref.  \onlinecite{land6} for the corresponding curve). 

We also note that the temperature dependence of $H_{c2}$ is quite different from predictions of the BCS theory (see Fig. 1).\cite{wert} This means that values of $H_{c2}(0)$ for HTSC's, evaluated from high-temperature $H_{c2}$ data using the corresponding formula of Ref. \onlinecite{wert}, are strongly overestimated (see also Ref. \onlinecite{sekitani}).

There are two main conclusions: (i) The Ginzburg-Landau parameter $\kappa$ is temperature independent. This follows from good agreement between the normalized $H_{c2}(T)$ curve, obtained by scaling of magnetization data, with direct measurements (see Fig. 1). (ii) $H_{c2}(T)/H_{c2}(0)$ is isotropic. This statement was initially made on the basis of the analysis of magnetization data collected on polycrystalline samples.\cite{land2} Now it is also confirmed by direct comparison of $H_{c2}(T)$ curves for two different orientations of the magnetic field (Fig. 1). 

While both conclusions were made independently, according to the Ginzburg-Landau theory, the second one follows from the first. Indeed, $H_{c2}(T) = \sqrt{2} \kappa H_c(T)$ where $H_c$ is the thermodynamic critical field, which cannot be anisotropic. Therefore, anisotropy of $H_{c2}$ may arise from the anisotropy of $\kappa$ only. If $\kappa$ does not depend on temperature, as it follows from the discussion above, $H_{c2}(T)/H_{c2}(0) = H_c(T)/H_c(0)$, i.e., the function $F$ in Eq. (2) is isotropic.  

Although direct measurements of $H_{c2}(T)$ are only available for YBa$_2$Cu$_3$O$_{7-x}$ samples, there cannot be much doubts that both conclusions are also valid for many other superconductors belonging exhibiting the same normalized $H_{c2}(T/T_c)$ curves.\cite{land1,land2,land3,land4}

In conclusion, it was demonstrated that temperature dependences of the normalized upper critical field, which were established by scaling of magnetization data collected in fields $H \ll H_{c2}$, are in very good agreement with recent direct measurements of $H_{c2}(T)$ in megagauss magnetic fields.\cite{miura,sekitani} This agreement shows  that the temperature dependence of the Ginzburg-Landau parameter in HTSC's is rather weak. Another result of the presented analysis is that $H_{c2}(T)/H_{c2}(0)$ is isotropic.

\begin{acknowledgments}

This work was in part supported by he NCCR MaNEP-II of the Swiss National Science Foundation (Project 4) and performed in the group of J. Hulliger.

\end{acknowledgments}


\begin{thebibliography}{71}
\bibitem{footn1} It is absolutely unclear why $T_c = 90.5$ K was used in Fig. 4(b) of Ref. \onlinecite{sekitani-pb}, in which data of Ref. \onlinecite{miura} are replotted as a function of $T/T_c$. As may be seen in Fig. 5 of Ref. \onlinecite{miura}, this value of $T_c$ is well above superconducting transition for this sample.
\end{thebibliography}
\end{document}